\documentclass[journal]{IEEEtran}

\usepackage{mathrsfs}
\usepackage{amsmath}
\usepackage{amsfonts}
\usepackage{amsthm}
\usepackage{amssymb}
\usepackage{stmaryrd}
\usepackage{graphicx}
\usepackage{subfigure}
\usepackage{indentfirst}
\usepackage{array}
\usepackage{cite}
\usepackage{enumerate}
\usepackage{bm}
\usepackage[linesnumbered,ruled,vlined]{algorithm2e}
\usepackage{algorithmic}
\usepackage{multirow}
\usepackage{epstopdf}
\usepackage{verbatim}
\usepackage{stfloats}
\usepackage{color}
\usepackage{booktabs}
\usepackage{footnote}

\SetKwComment{Comment}{//}{}

\begin{document}

\title{\LARGE{Joint Successive Cancellation List Decoding for the Double Polar codes}}

\author{Yanfei~Dong,
        Kai~Niu,~Jincheng~Dai,~Sen~Wang,~and~Yifei Yuan%
\thanks{This work was supported by the National Key R\&D Program of China (No. 2018YFE0205501), National Natural Science Foundation of China (No. 62071058 \& No. 62001049), and Beijing University of Posts
and Telecommunications - China Mobile Research Institute Joint Innovation
Center.}
\thanks{The authors are with the Key Laboratory of Universal Wireless Communications, Ministry of Education, Beijing University of Posts and Telecommunications, Beijing 100876, China (email: niukai@bupt.edu.cn).}
\vspace{-2em}
}

\maketitle

\begin{abstract}
As a new joint source-channel coding scheme, the double polar (D-Polar) codes have been proposed recently.
In this letter, a novel joint source-channel decoder, namely the joint successive cancellation list (J-SCL) decoder, is proposed to improve the decoding performance of the D-Polar codes.
We merge the trellis of the source polar code and that of the channel polar code to construct a compound trellis.
In this compound trellis, the joint  source-channel nodes represent both of the information bits and the high-entropy bits.
Based on the compound trellis, the J-SCL decoder is designed to recover the source messages by combining the source SCL decoding and channel SCL decoding.
The J-SCL decoder doubles the number of the decoding paths at each decoding level and then reserves the $L$ paths with the smallest joint path-metric (JPM).
For the JSC node, the JPM is updated considering both the channel decision log-likelihood ratios (LLRs) and the source decision LLRs.
Simulation results show that the J-SCL decoder outperforms the turbo-like BP (TL-BP) decoder with lower complexity.
\end{abstract}

\begin{IEEEkeywords}
Double polar codes, joint source-channel decoder, joint successive cancellation list decoder.
\end{IEEEkeywords}

\IEEEpeerreviewmaketitle

\section{Introduction}
\IEEEPARstart{P}{olar} codes\cite{Arikan2009}, invented by Ar$\i$kan, are the first theoretically provable capacity-achieving error correction codes for any binary-input discrete memoryless channels with low encoding and decoding complexity.
The successive cancellation (SC) \cite{Arikan2009} and belief propagation (BP) decoders \cite{Arkan2008} are two commonly decoding methods of polar codes.
However, the finite-length performance of polar codes is unsatisfying under the SC and BP decoders.
The successive cancellation list (SCL) decoder \cite{Chen2012,Niu2012,Tal2015,BalatsoukasStimming2015} has been proposed to improve the performance of polar codes.
Moreover, the source polarization has been introduced and investigated in \cite{Arikan2010} as the complement of channel polarization.
Polar codes are proved to be optimal for lossy source coding \cite{Korada2010}, which can achieve the rate-distortion bound for a binary symmetric source.
For lossless compression, a polar encoding algorithm achieving the optimal compression rate asymptotically is developed in \cite{Cronie2010}.

Polar codes have been used in the field of joint source and channel coding (JSCC).
By modeling the source redundancy as a sequence of $t$-erasure correcting block codes, polar codes have proven to be a good candidate to exploit the benefit of source redundancy in a JSCD scheme \cite{aaJiangISIT}.
For correlated sources, a joint decoding scheme using systematic polar codes is proposed in \cite{Jin2018}, which exploits the correlation among the sources.
In \cite{Jin2018a}, a JSCC scheme using quasi-uniform systematic polar code is proposed by introducing additional bit-swap coding to modify original polar coding.
Inspired by the double low-density parity-check (LDPC) codes \cite{Fresia2010,JiguangHe2012,Liu2020}, a JSCC scheme using double polar (D-Polar) codes is proposed in our previous work \cite{Dong2021}, in which the source is first compressed by a source polar code, and then the compressed bits are protected by a systematic polar code (SPC).
And a turbo-like BP (TL-BP) decoder is also designed for the decoding of D-Polar codes, which suffers from severe error floor and high complexity.

In this letter, we propose a joint successive cancellation (J-SCL) decoding algorithm to improve the performance of the D-Polar codes.
In the D-Polar codes, the source is first compressed by a polar code, and then another polar code is employed to protect the compressed source.
We find that the information bit part of channel polar code is a copy of the high entropy bit part of source polar code.
Therefore, a compound trellis can be constructed by using the joint source-channel (JSC) nodes to represent both of the information bits and the high-entropy bits, merging the trellis of the source polar code and that of the channel polar code.
In this combined trellis, the variable nodes corresponding to frozen bits and low-entropy bits are referred to as frozen nodes and low-entropy nodes, respectively.

Based on the compound trellis, a JSCD scheme, namely the J-SCL decoder, is proposed, which doubles the number of paths at each JSC node and low-entropy node.
When the number of decoding paths exceeds the \emph{list size} $L$, a pruning procedure is used to select the $L$ candidate paths with the smallest joint path-metric (JPM) from the list.
For the JSC node, the update of the JPM considers both the channel decision log-likelihood ratios (LLRs) and the source decision LLRs.
For the low-entropy node, the update of the JPM considers only the source decision LLRs.
Simulation results show that the J-SCL decoder gains 0.42 dB compared to the TL-BP decoder, while the error floor of bit error rate (BER) is lowered from $10^{-4}$ to $10^{-6}$.


\section{Joint Successive Cancellation List Algorithm}
In this section, we first introduce the D-Polar JSCC system model.
Secondly, we give the method to fuse the trellis of the source polar code and that of channel polar code into a compound trellis.
Finally, we propose the J-SCL decoder for decoding the D-Polar codes.

\subsection{Notational Conventions}
In this letter, the calligraphic characters, such as $\mathcal{X}$, are used to denote sets.
We write the notation $a^N_1$ to denote a row vector $(a_1,\ldots,a_N)$.
Given such a vector $a^N_1$, we use $a^j_i$, $1\leq i$, $j\leq N$, to denote the subvector $(a_i,\ldots,a_j)$, and $a_i$ denotes the $i$-th element in $a^N_1$.
For positive integer $N$, $\llbracket{N}\rrbracket\triangleq\{1,2,\ldots,N\}$.
Given $\mathcal{A}\subset\llbracket{N}\rrbracket$, we write $a_\mathcal{A}$ to denote the subvector $(a_i:i\in\mathcal{A})$ of $a^N_1$.
\subsection{System Model}
\begin{figure}
\centering
\includegraphics[scale=0.5]{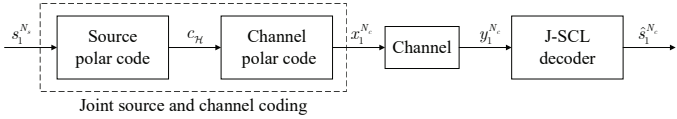}
\caption{The D-Polar codes JSCC system framework.}
\label{Fig1}
\end{figure}
Fig. \ref{Fig1} illustrates the framework of the D-Polar codes JSCC system framework.
Let $S\sim \textrm{Ber}(p)$ denote a binary Bernoulli random variable and $P_S(1)=p$.
The source polar code can be identified by a parameter vector $(N_s,K,\mathcal{H})$, where $N_s=2^{n_s}$ is the length of the source vector, and $K$ specifies the size of the \emph{high-entropy set} $\mathcal{H}$ \cite{Arikan2010}.
The ratio $K/N_s$ is called the compression rate $R_s$.
The sequence of $N_s$ i.i.d. source symbols $s^{N_s}_1$  generated from $S$ is first compressed into $c_{\mathcal{H}}$ by using a source polar code.
In the source encoding process, we compute $c^{N_s}_1=s^{N_s}_1G_{N_s}$ and output $c_\mathcal{H}$ as the compressed word, where $G_{N_s}=F^{\otimes n_s}$ equals the $n_s$-th Kronecker power of the matrix
$F\triangleq
\left[\begin{smallmatrix}
1&0\\
1&1
\end{smallmatrix}\right]$.

Second, the compressed word $c_\mathcal{H}$ is protected by another polar code.
The channel polar code can be identified by a parameter vector $(N_c,K,\mathcal{A},u_{\mathcal{A}^c})$, where $N_c=2^{n_c}$ is the length of a codeword, $K$ specifies the size of the \emph{information set} $\mathcal{A}$ and $u_{\mathcal{A}^c}$ denotes $\emph{frozen}$ bits\cite{Arikan2009}.
The ratio $K/N_c$ is called the channel coding rate $R_c$ and the overall rate of the D-Polar codes JSCC system is $R=R_c/R_s$.
For the channel polar code, the compressed word $c_\mathcal{H}$ is transmitted over the information bits $u_\mathcal{A}$.
The channel polar coding is executed by
\begin{equation}\label{chPolarCoding}
\begin{split}
x^{N_c}_1&=u^{N_c}_1G_{N_c}\\
&=u_\mathcal{A}G_{N_c}(\mathcal{A})\oplus u_{\mathcal{A}^c}G_{N_c}(\mathcal{A}^c)\\
&=c_\mathcal{H}G_{N_c}(\mathcal{A})\oplus u_{\mathcal{A}^c}G_{N_c}(\mathcal{A}^c)\\
\end{split}
\end{equation}
where $G_{N_c}=F^{\otimes n_c}$ and $G_{N_c}(\mathcal{A})$ denotes the submatrix of $G_{N_c}$ formed by the rows with indices in $\mathcal{A}$.
The codeword $x^{N_c}_1$ is then transmitted through a channel.
Having received $y^{N_c}_1$ and known the $P_S(1)=p$, the proposed J-SCL decoder will build an estimate $\hat{s}^{N_s}_1$ of $s^{N_s}_1$.
\subsection{Source-Channel Trellis Merging}
The trellis of the source polar code and that of the channel polar code can be merged into a compound trellis.
From (\ref{chPolarCoding}), we have known $u_\mathcal{A}=c_\mathcal{H}$, which indicates that the information bits are duplicates of the high-entropy bits.
The merging of the source trellis and the channel trellis is based on the fusion of the high-entropy bits of the source polar code and the information bits of the channel polar code.

\begin{figure}
\centering
\includegraphics[scale=0.46]{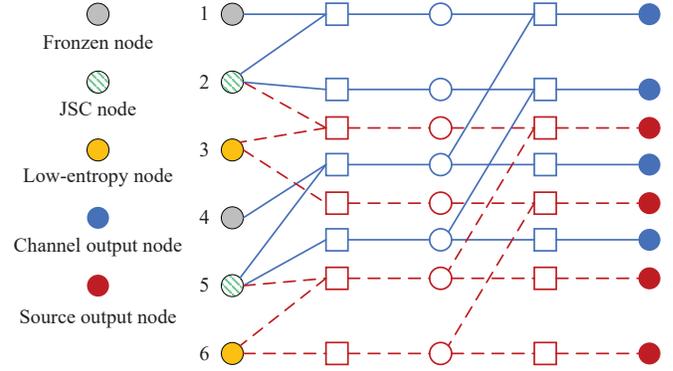}
\caption{An example of the compound trellis.}
\label{Fig2}
\end{figure}
For the sake of understanding, we give an example of  the compound trellis, as shown in Fig. \ref{Fig2}.
The source polar code is identified by $(N_{s}=4, K=2, \mathcal{H}=\{1,3\})$ and the channel polar code is indicated by $(N_{c}=4, K=2, \mathcal{A}=\{2,4\})$.
In Fig. \ref{Fig2}, the channel polar code is represented by a blue trellis, and the source polar code is described by a red trellis.
In this compound trellis, the leftmost variable nodes can be divided into three types of nodes, i.e., frozen nodes, JSC nodes, and low-entropy nodes.
The frozen node and the low-entropy node correspond to the frozen bit of the channel polar code and the low-entropy bit of the source polar code, respectively.
And the JSC node corresponds to the information bit of the channel polar code as well as the high-entropy bit of the source polar code.
Let $\mathcal{J}$ denote the index set of JSC nodes and $\mathcal{W}$ denote the index set of low-entropy nodes.
For this compound trellis, we have $\mathcal{J}=\{2,5\}$, $\mathcal{W}=\{3,6\}$, and $N=6$,  where $N=N_s+N_c-K$ denotes the number of the leftmost variables.

The D-Polar codes can be decoded as a compound trellis using the J-SCL decoder.
For polar codes, the sequential decoding schedule of an SCL decoder can be determined by the information set $\mathcal{A}$.
Correspondingly, the sets $\mathcal{J}$ and $\mathcal{W}$ determine the sequential decoding schedule of the J-SCL decoder.

\begin{algorithm}
\setlength{\abovecaptionskip}{0.cm}
\setlength{\belowcaptionskip}{-0.cm}
\caption{The construction of the set $\mathcal{J}$ and $\mathcal{W}$}
\label{JSC-PC}
\KwIn{$N_s, N_c, K, \mathcal{H}, \mathcal{A}$}
\KwOut{$\mathcal{J}$, $\mathcal{W}$}
\For {$i=1,2,\ldots,K$}
{
    $\epsilon=\sum^i_{k=1}(h_k-h_{k-1}-1)$;\\
    $j_i\leftarrow a_i+\epsilon$;\\
}
Set $\mathcal{H}^c\leftarrow \llbracket{N_s}\rrbracket\setminus \mathcal{H}$;\\
\For {$i=1,2,\ldots,N_s-K$}
{
    $\varepsilon=\sum^i_{k=1}(h^c_k-h^c_{k-1}-1)$;\\
    $\tau=\sum^\varepsilon_{k=1}(a_k-a_{k-1}-1)$;\\
    $w_i\leftarrow h^c_i+\tau$;\\
}
\Return $\mathcal{J}$, $\mathcal{W}$.
\end{algorithm}
Given the source polar code and the channel polar code, Algorithm \ref{JSC-PC} shows the construction method of the set $\mathcal{J}$ and $\mathcal{W}$.
Let $a_i$, $h_i$, and $j_i$ denote the $i$-th element of $\mathcal{A}$, $\mathcal{H}$, and $\mathcal{J}$, respectively.
For the leftmost variables of the channel trellis, the construction of the compound trellis is equivalent to inserting the low-entropy node into the channel trellis.
The number of low-entropy bits before the high-entropy bit $c_{h_i}$ can be calculated by $\epsilon=\sum^i_{k=1}(h_k-h_{k-1}-1)$, where $h_0$ is set to 0.
As shown in Line 3 of Algorithm \ref{JSC-PC}, we can obtain $j_i$ by adding $\epsilon$ to $a_i$.

Similarly, the construction of the compound trellis is equivalent to the insertion of the frozen node into the source trellis.
The set $\mathcal{H}^c$ is the complement of the high-entropy set $\mathcal{H}$, which is called the low-entropy set.
The $i$-th element of $\mathcal{H}^c$ is denoted by $h^c_{i}$.
The number of high-entropy bits before the low-entropy $c_{h^c_{i}}$ can be computed by $\varepsilon=\sum^i_{k=1}(h^c_k-h^c_{k-1}-1)$, where $h^c_0$ is set to 0.
In the D-Polar codes, the high-entropy bit $c_{h_\varepsilon}$ is carried by the information bit $u_{a_\varepsilon}$.
The number of frozen bits between $u_{a_i}$ and $u_{a_{i-1}}$ can be obtained by $a_i-a_{i-1}-1$, where $a_0$ is set to 0.
Then, we can calculate the number of frozen bits before the information bit $u_{a_\varepsilon}$ by $\tau=\sum^\varepsilon_{k=1}(a_k-a_{k-1}-1)$.
As shown in Line 7 of Algorithm \ref{JSC-PC}, the $i$-th element of $\mathcal{W}$ is denoted by $w_i$, which can be obtained by $h^c_i+\tau$.
Finally, we can get the set $\mathcal{J}$ and $\mathcal{W}$.

\subsection{Joint Successive Cancellation List Decoding}
A J-SCL decoder is proposed for the joint source and channel decoding of the D-Polar codes.
Algorithm \ref{J-SCL} shows a description of the J-SCL decoder.
The J-SCL decoding algorithm is a breadth-first search with $N$ levels, which is governed by the list size $L$.
The bits corresponding to the JSC nodes and low-entropy nodes are decoded successively one-by-one.
At each decoding stage, $L$ decoding paths are considered concurrently.
Let $\mathcal{L}$ represent the path list of the J-SCL decoder, and $v^\varphi_1$ denote a candidate path $\ell$  for $1\leq\varphi\leq N$.
The J-SCL decoder doubles the number of paths at each JSC node and low-entropy node.
\begin{algorithm}
\setlength{\abovecaptionskip}{0.cm}
\setlength{\belowcaptionskip}{-0.cm}
\caption{J-SCL Decoder}
\label{J-SCL}
\KwIn{$N, \mathcal{J}$, $\mathcal{W}$, $p$, $y^{N_c}_1$}
\KwOut{$\hat{s}^{N_s}_1$}
Initialize $\mathcal{L}\leftarrow \{0\}$, $\textrm{JPM}^{(0)}_0\leftarrow 0$, $i_c\leftarrow0$, $i_s\leftarrow0$;\\
\For{$\varphi=1,2,\ldots,N$}
{
    \For{$\ell\in\mathcal{L}$}
    {
        \If{$\varphi\in \mathcal{J}$}
        {
            Compute $L^{(i_c)}_{N_c}[\ell]$ and $L^{(i_s)}_{N_s}[\ell]$;\\
            Copy the path $\ell$ into a new path $\ell^\prime\notin\mathcal{L}$ ;\\
            $(\hat{v}_\varphi[\ell], \mathrm{JPM}^{\varphi}_\ell)\leftarrow(0,\tilde{\phi}(\textrm{JPM}^{(\varphi-1)}_\ell, \textsf{L}^{(i_c)}_{n_c}[\ell], \textsf{L}^{(i_s)}_{n_s}[\ell],0))$;\\
            $(\hat{v}_\varphi[\ell^\prime], \mathrm{JPM}^{\varphi}_{\ell^\prime})\leftarrow(1,\tilde{\phi}(\textrm{JPM}^{(\varphi-1)}_\ell, \textsf{L}^{(i_c)}_{n_c}[\ell], \textsf{L}^{(i_s)}_{n_s}[\ell],1))$;\\
             $i_c\leftarrow i_c+1$, $i_s\leftarrow i_s+1$;\\
        }
        \ElseIf{$\varphi\in \mathcal{W}$}
        {
            Compute $L^{(i_s)}_{N_s}[\ell]$;\\
            Copy the path $\ell$ into a new path $\ell^\prime\notin\mathcal{L}$ ;\\
            $(\hat{v}_\varphi[\ell], \mathrm{JPM}^{\varphi}_\ell)\leftarrow(0,\phi(\textrm{JPM}^{(\varphi-1)}_\ell, \textsf{L}^{(i_s)}_{n_s}[\ell],0))$;\\
            $(\hat{v}_\varphi[\ell^\prime], \mathrm{JPM}^{\varphi}_{\ell^\prime})\leftarrow(1,\phi(\textrm{JPM}^{(\varphi-1)}_\ell, \textsf{L}^{(i_s)}_{n_s}[\ell],1))$;\\
            $i_s\leftarrow i_s+1$;\\
        }
        \Else
        {
            Compute $L^{(i_c)}_{N_c}[\ell]$;\\
            $(\hat{v}_\varphi[\ell], \mathrm{JPM}^{\varphi}_\ell)\leftarrow(v_\varphi,\phi(\textrm{JPM}^{(\varphi-1)}_\ell, \textsf{L}^{(i_s)}_{n_s}[\ell],v_\varphi))$;\\
            $i_c\leftarrow i_c+1$;\\
        }
    }
    Discard all but $L$ paths with the smallest $\textrm{JPM}^{(i)}_\ell$;\\
}
$\ell^*\leftarrow \textrm{arg}\min_{\ell\in\mathcal{L}}\textrm{JPM}^{(N)}_\ell$;\\
$\hat{s}^{N_s}_1\leftarrow \hat{v}_{\mathcal{\bar{A}}}[\ell^*]G_{N_s}$\\
\Return $\hat{s}^{N_s}_1$.
\end{algorithm}
At each level $\varphi\in\mathcal{J}\cup\mathcal{W}$ of the path $\ell$, the path $\hat{v}^{\varphi-1}_1$ is split into two paths, e.g., $\ell$ and $\ell^\prime$.
Both of the path $\ell$ and $\ell^\prime$ have $\hat{v}^{\varphi-1}_1$ as a prefix, and the path $\ell$ ends with ``0" while the other path $\ell^\prime$ ends at ``1".
For a path $\ell$ at level $\varphi$, let the \emph{joint path-metric} be denoted as $\textrm{JPM}^{(\varphi)}_\ell$.
As soon as the number of paths exceeds $L$, $L$ paths with the smallest $\textrm{JPM}^{(\varphi)}_\ell$ are reserved (Line 20, Algorithm \ref{J-SCL}).
At the end of decoding, the decoding path with the smallest $\textrm{JPM}^{(N)}_\ell$ is selected as the final decoding path.
\subsubsection{JSC node decoding}
The decoding of the JSC node is shown in Lines 5-9 of Algorithm \ref{J-SCL}.
The level $\varphi\in\mathcal{J}$ means that the bit $v_{\varphi}$ correspond to the JSC node.
Therefore, the update of $\textrm{JPM}^{(\varphi)}_\ell$ need to consider the channel decision log-likelihood ratios (LLRs) $\textsf{L}^{(i_c)}_{n_c}[\ell]$ and the source decision LLRs $\textsf{L}^{(i_s)}_{n_s}[\ell]$ together,
where $i_c$ and $i_s$ denote the current decoding levels for the channel SC decoding and source SC decoding corresponding to the level $\varphi$.
The decision LLRs $\textsf{L}^{(i_c)}_{n_c}[\ell]$ and $\textsf{L}^{(i_s)}_{n_s}[\ell]$ can be computed by the recursions \cite{Arikan2009,BalatsoukasStimming2015}
\begin{gather*}
\textsf{L}^{(2i)}_n[\ell] = f(\textsf{L}^{2i-[i \bmod 2^{n-1}]}_{n-1}[\ell],\textsf{L}^{2^n+2i-[i \bmod 2^{n-1}]}_{n-1}[\ell]),\\
\textsf{L}^{(2i+1)}_n[\ell] = g(\textsf{L}^{2i-[i \bmod 2^{n-1}]}_{n-1}[\ell],\textsf{L}^{2^n+2i-[i \bmod 2^{n-1}]}_{n-1}[\ell],\textsf{u}^{2i}_n[\ell]),
\end{gather*}
where the operation $f$ and $g$ are defined as
\begin{gather}
f(\alpha, \beta) \triangleq \ln \left(\frac{e^{\alpha+\beta}+1}{e^{\alpha}+e^{\beta}}\right), \\
g(\alpha, \beta, u) \triangleq(-1)^{\textsf{u}} \alpha+\beta,
\end{gather}
respectively. For the channel SC decoding, the recursions terminate at
\begin{equation}
\textsf{L}_{0}^{(i)}[\ell] \triangleq \ln\frac{W(y_i|0)}{W(y_i|1)}, \quad \forall i \in \llbracket N_c \rrbracket,
\end{equation}
where $W(y|x)$ is the probability distribution function of the output letter $y$ when $x$ is transmitted.
Correspondingly, for the source SC decoding, we have
\begin{equation}
\textsf{L}_{0}^{(i)}[\ell] \triangleq \ln\frac{1-P_S(1)}{P_S(1)}, \quad \forall i \in \llbracket N_s \rrbracket.
\end{equation}
The calculation of \emph{partial sums} is written as
\begin{equation}
\begin{aligned}
\textsf{u}_{n-1}^{\left(2 i-\left[i \bmod 2^{s-1}\right]\right)}[\ell] &=\textsf{u}_{n}^{(2 i)}[\ell] \oplus \textsf{u}_{n}^{(2 i+1)}[\ell], \\
\textsf{u}_{n-1}^{\left(2^{s}+2 i-\left[i \bmod 2^{s-1}\right]\right)}[\ell] &=\textsf{u}_{n}^{(2 i+1)}[\ell],
\end{aligned}
\end{equation}
which is computed starting from $\textsf{u}^{(i)}_n[\ell]\triangleq\hat{v}_\varphi[\ell]$.
Since the JSC nodes connect both the trellis of channel polar code and that of source polar code, the update of the $\textrm{JPM}^{(\varphi)}_\ell$ needs to merge the channel decision LLRs $\textsf{L}^{(i_c)}_{n_c}[\ell]$ and the source decision LLRs $\textsf{L}^{(i_s)}_{n_s}[\ell]$.
When the decision LLRs $\textsf{L}^{(i_c)}_{n_c}[\ell]$ and $\textsf{L}^{(i_s)}_{n_s}[\ell]$ are obtained, for the bits $\{v_\varphi|\varphi\in\mathcal{J}\}$, the update of $\textrm{JPM}^{(\varphi)}_\ell$ is written as
\begin{equation}
\textrm{JPM}^{(\varphi)}_{\ell}=\tilde{\phi}(\textrm{JPM}^{(\varphi-1)}_\ell, \textsf{L}^{(i_c)}_{n_c}[\ell], \textsf{L}^{(i_s)}_{n_s}[\ell],\hat{v}_\varphi[\ell]),
\end{equation}
where the function $\tilde{\phi}$ is defined as
\begin{equation}
\label{JSC_JPM}
\tilde{\phi}(\mu,\lambda_c,\lambda_s,v)\triangleq\mu+\ln(1+e^{(2v-1)\lambda_c})+\ln(1+e^{(2v-1)\lambda_s})
\end{equation}
\subsubsection{Low-entropy node decoding}
The decoding operation of the low-entropy node is shown in Lines 11-15 of Algorithm \ref{J-SCL}.
Similar to the level $\varphi\in\mathcal{J}$, we copy the path $\ell$ into a new path $\ell^\prime\notin\mathcal{L}$ at the level $\varphi\in\mathcal{W}$.
The path $\ell$ and $\ell^\prime$ have the same prefix, e.g., $\hat{v}^{\varphi-1}_1[\ell]$.
The last bit of the path $\ell$ is set to ``0", and the last bit of path $\ell$ is set to ``1".
Since $v_\varphi$ correspond to a low-entropy node, the update of $\textrm{JPM}^{(\varphi)}_\ell$ only need to consider the source decision LLR $\textsf{L}^{(i_s)}_n$, which can be written as
\begin{equation}
\textrm{JPM}^{(\varphi)}_{\ell}=\phi(\textrm{JPM}^{(\varphi-1)}_\ell, \textsf{L}^{(i_s)}_{n}[\ell], \hat{v}_\varphi[\ell]).
\end{equation}
The function $\phi(\mu,\lambda,v)$ is defined as
\begin{equation}
\phi(\mu,\lambda,v)\triangleq\mu+\ln(1+e^{(2v-1)\lambda}).
\end{equation}

\subsubsection{Frozen node decoding}
The decoding operation of the frozen node is shown in Lines 17-19 of Algorithm \ref{J-SCL}.
Let $\bar{\mathcal{A}}=\mathcal{J}\cup\mathcal{W}$.
The subvector $v_{\bar{\mathcal{A}}^c}$ equals the frozen bits $u_{\mathcal{A}^c}$, which is known to both the transmitter and the receiver.
Therefore, the $\hat{v}_\varphi[\ell]$ can be directly set to $v_{\varphi}$ at each level $\varphi\notin(\mathcal{J}\cup\mathcal{W})$.
The update of $\textrm{JPM}^{(\varphi)}_\ell$ only needs to consider the channel decision LLR $\textsf{L}^{(i_c)}_n$, which can be written as
\begin{equation}
\textrm{JPM}^{(\varphi)}_{\ell}=\phi(\textrm{JPM}^{(\varphi-1)}_\ell, \textsf{L}^{(i_c)}_{n}[\ell], v_\varphi).
\end{equation}

When the decoding level $\varphi$ reaches $N$, the path $\ell$ with the smallest $\textrm{JPM}^{(N)}_\ell$ is selected as the final decoding path $\ell^*$.
The estimate $\hat{c}^{N_s}_1$ of $c^{N_s}_1$ equals $\hat{v}_{\bar{\mathcal{A}}}[\ell^*]$, and the source decision $\hat{s}^{N_s}_1$ can be calculated by
\begin{equation}
\hat{s}^{N_s}_1= \hat{v}_{\mathcal{\bar{A}}}[\ell^*]G_{N_s}
\end{equation}
which is shown in Line 21 Algorithm \ref{J-SCL}.

\section{Performance Evaluation}
In this section, we provide simulation results and analysis of the proposed J-SCL decoder over the
binary phase shift keying (BPSK) modulated additive white Gaussian noise (AWGN) channel.
The polar code used in simulations is constructed via Gaussian approximation \cite{Trifonov2012}.
The length of source vector is $N_s=512$, and the overall rate $R$ is set to $1/2$.
The maximum of simulated frame number is $10^8$, and the simulation terminates when the error frame number reaches $10^2$ at each $E_b/N_0$.
Note that $E_b$ refers to energy per source bit.
\subsection{Simulation Results}
\begin{figure}
\centering
\includegraphics[scale=0.63]{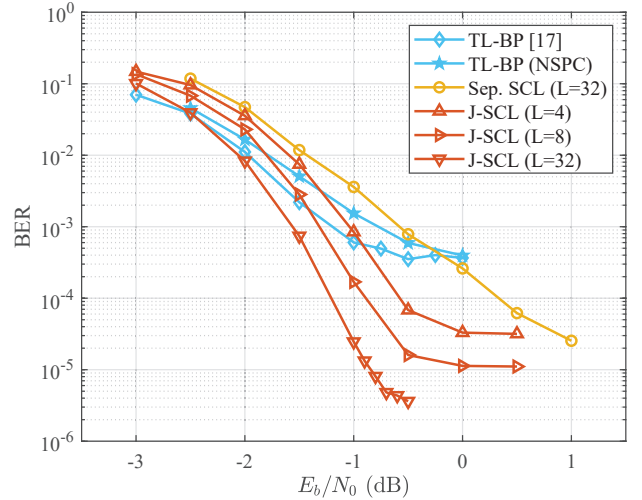}
\caption{BER performance of TL-BP decoder, Separate SCL decoder and J-SCL decoder.}
\label{Fig3}
\end{figure}
 \begin{figure}
\centering
\includegraphics[scale=0.77]{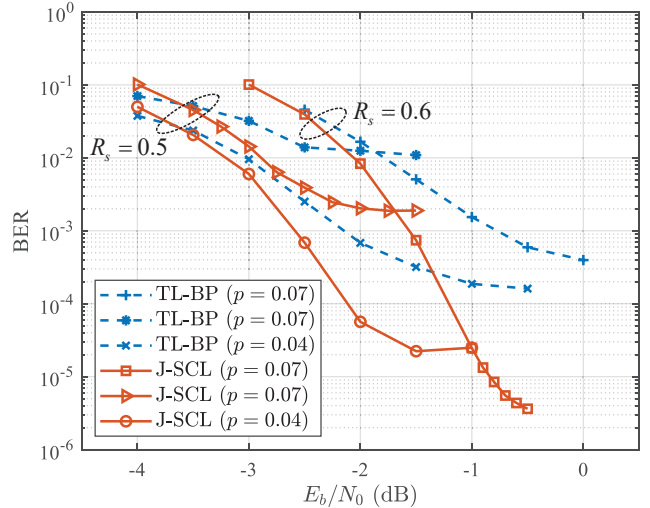}
\caption{BER performance of J-SCL $(L=32)$ decoding and TL-BP decoding with various $P_S(1)=p$ and the compression rate $R_s$.}
\label{Fig4}
\end{figure}
In Fig. \ref{Fig3}, we compare the BER performance of various decoding schemes for D-Polar codes with $p=0.07$ and $R_s=0.6$.
The TL-BP decoder can be referred to \cite{Dong2021}.
And the Separate SCL (Sep. SCL) decoder refers to a channel SCL decoder followed by an independent source SCL decoder.
From this figure, we can see that the performance gain of J-SCL ($L=4$) decoding versus Sep. SCL ($L=32$) decoding can reach 0.96 dB at $\text{BER}=10^{-4}$.
When the list size $L$ increases to 8 and 32, the gain of J-SCL decoding compared to the Sep. SCL ($L=32$) decoding increases to 1.22 dB and 1.54 dB at $\text{BER}=10^{-4}$, respectively.
The gain of J-SCL decoding with respect to Sep. SCL decoding is due to the residual redundancy of the source code considered in the channel decoding.
The residual redundancy that is not removed by the source polar encoder gives additional information to the channel SCL decoding.
For the Sep. SCL decoder, the independent channel SCL decoder cannot benefit from the residual redundancy.
Compared with existing JSCD scheme of D-Polar codes, i.e., TL-BP decoder \cite{Dong2021}, the J-SCL ($L=32$) can obtain 0.36 dB gain at $\text{BER}=10^{-3}.$
One of the weaknesses of the TL-BP decoder is the high error floor, which is due to the residual decoding errors introduced by the limited length of source code.
In contrast, a significant advantage of the J-SCL ($L=32$) is the reduction of the error floor from $10^{-4}$ to $10^{-6}$.
The list size $L$ affects the error floor of the J-SCL decoding.
When the list size is reduced to 8 and 4, the error floor will be higher than $10^{-5}$.
The lower error floor of the J-SCL decoder is due to the better performance of source decoding using SCL decoder than BP decoder.
Moreover, this figure gives the impact on the performance of the TL-BP decoder caused by replacing the SPC in D-Polar codes with the non-systematic polar code (NSPC).
The TL-BP decoder for D-Polar codes using NSPC has 0.42 performance loss compared with the TL-BP decoder for original D-Polar codes \cite{Dong2021} at $\text{BER}=10^{-3}$, while the error floor is close.
 The performance gain of J-SCL ($L=32$) decoding versus TL-BP (NSPC) decoding can reach 0.78 dB at $\text{BER}=10^{-3}$.

 In Fig. \ref{Fig4}, we compare the BER performance of J-SCL decoding and TL-BP decoding at different $p$ and $R_s$.
 In this simulation, the list size $L$ is set to 32.
 The J-SCL decoder and TL-BP decoder use the same transmitter.
 From this figure, we can see that the J-SCL outperforms the TL-BP decoder with various $p$ and $R_s$.
The performance gap between the J-SCL decoding and TL-BP decoding widens as the $E_b/N_0$ increases.
For the J-SCL decoder with $p=0.07$, the performance gain of $R_s=0.5$ versus $R_s=0.6$ reaches 0.84 dB.
Given a constant overall rate $R$, this additional gain is due to the lower channel coding rate $R_c$.
 However, the compression rate of $R_s=0.5$ causes the error floor to rise to $10^{-3}$.
 The compression rate $R_s$ trades off the coding gain and the error floor.
 Moreover, this figure shows that a lower entropy will result in a better BER performance.
 For the J-SCL decoder with $R_s=0.5$, the performance gain of $p=0.04$ versus $p=0.07$ reaches 0.32 dB, and the error floor falls from $10^{-3}$ to $10^{-5}$.
 \begin{figure}
\centering
\includegraphics[scale=0.63]{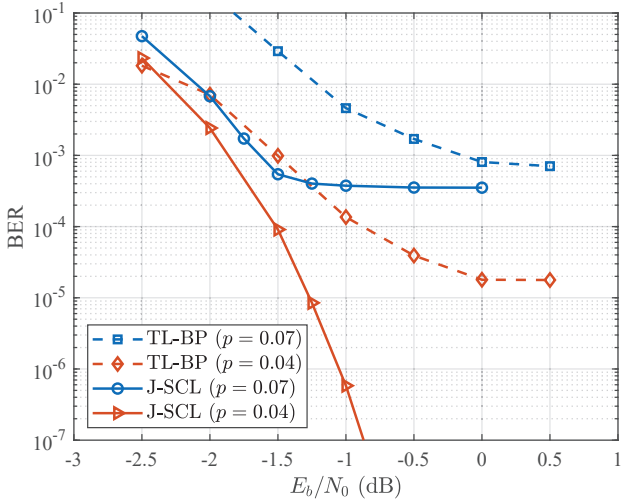}
\caption{BER performance of J-SCL $(L=32)$ decoding and TL-BP decoding with the overall rate $R=1$.}
\label{Fig5}
\end{figure}
Moreover, Fig. \ref{Fig5} provides a comparison of the J-SCL (L=32) decoder and the TL-BP decoder with the overall rate $R=1$.
The length of source vector is increased to 1024, and $R_s$ ($R_c$) is set to $1/2$.
For $p=0.07$, the J-SCL decoder achieves 1.47 dB gain compared with the TL-BP decoder at $\text{BER}=10^{-3}$.
For $p=0.04$, the J-SC decoder outperforms the TL-BP decoder by about 0.64 dB at $\text{BER}=10^{-4}$.
Compared with the error floor of the TL-BP decoder at $10^{-5}$, the BER curve of the J-SCL decoder remains without an error floor at $10^{-7}$.
\subsection{Complexity Analysis}
Recall Section II, we can see that the calculation of $L^{(i_c)}_{N_c}[\ell]$ involves a channel SC decoding, and the calculation of $L^{(i_s)}_{N_s}[\ell]$ involves a source SC decoding.
Therefore, the complexity of J-SCL decoding is $O(LN_c\log N_c+LN_s\log N_s)$.
As shown in \cite{Dong2021}, the TL-BP decoder contains a channel BP decoder and a source BP decoder.
Let $I$ denote the maximum number of iterations of TL-BP.
Then, the complexity of the TL-BP decoder can be represented by $O(IN_c\log N_c+IN_s\log N_s)$.
Although the TL-BP decoder can use the early stopping criterion to obtain a very low average number of iterations in the high $E_b/N_0$ region, the iteration number is still $I$ in the worst-case.
The J-SCL decoder outperforms  in terms of complexity as long as $I$ is greater than $L$.

\section{Conclusion}
In this letter, we propose a J-SCL decoder for the D-Polar codes JSCC scheme.
To facilitate the design of the J-SCL decoder, we replaced the SPC in the D-Polar codes with an NSPC.
Then, we give a method to combine the trellis of the channel polar code and that of the source polar code into a compound trellis, where a JSC node represents both the information bit and the high-entropy bit.
Based on the compound trellis, the J-SCL decoder is proposed by performing path splitting on the JSC nodes and the low-entropy nodes.
Simulation results show that the proposed JSCD outperforms the TL-BP decoder.

\ifCLASSOPTIONcaptionsoff
  \newpage
\fi

\bibliographystyle{IEEEtran}
\bibliography{bibfile}

\end{document}